\documentclass[letter]{aa} 
\usepackage{natbib}
\bibpunct{(}{)}{;}{a}{}{,} 
\usepackage{graphicx}
\usepackage{epstopdf}
\usepackage{txfonts}
\begin{document}
   \title{Circumstellar effects on the Rb abundances in O-rich AGB stars}


   \author{O. Zamora
          \inst{1,2}
          \and
          D.~A. Garc{\'{\i}}a-Hern{\'a}ndez \inst{1,2}
          \and
          B. Plez
          \inst{3}
          \and
          A. Manchado \inst{1,2,4}
          }

   \institute{Instituto de Astrof{\'{\i}}sica de Canarias (IAC),
              E-38200 La Laguna, Tenerife, Spain\\
              \email{ozamora@iac.es}
        \and Departamento de Astrof{\'{\i}}sica, Universidad de La Laguna (ULL), E-38206 La Laguna, Tenerife, Spain
         \and
             Laboratoire Univers et Particules de Montpellier, Universit\'e Montpellier 2, CNRS, F-34095 Montpellier, France
          \and
             Consejo Superior de Investigaciones Cient\'{\i}ficas (CSIC), Spain \\ 
                          }

   \date{Received January xx, 2014; accepted xxxx xx, 2014}

 
\abstract {For the first time we explore the circumstellar effects on the Rb (and Zr)
abundance determination in O-rich asymptotic giant branch (AGB) stars by considering
the presence of a gaseous circumstellar envelope with a radial wind.
A modified version of the spectral synthesis code Turbospectrum was
used to deal with extended atmosphere models and velocity fields. The Rb and Zr
abundances were determined from the resonant 7800 \AA\ Rb I line and the  6474
\AA\ ZrO bandhead, respectively, in five representative O-rich AGB stars 
with different expansion velocities and metallicities. By using our new dynamical
models, the Rb I line profile (photospheric and circumstellar components) is
very well reproduced. Interestingly, the derived Rb abundances are much lower
(by 1-2 dex) in those O-rich AGB stars showing the higher circumstellar
expansion velocities. The Zr abundances, however, remain close to the solar
values. The Rb abundances and [Rb/Zr] ratios derived here significantly
resolve the problem of the present mismatch between the observations of
intermediate-mass (4--8 M$_{\sun}$) Rb-rich AGB stars and the AGB
nucleosynthesis theoretical predictions.}

   
   \maketitle
%

\section{Introduction}

The asymptotic giant branch (AGB) is the last nuclear-burning phase of low- and
intermediate-mass stars (0.8 $\leq$ M $\leq$ 8 M$_{\sun}$). The AGB stars are among
the main contributors to the chemical enrichment of the interstellar medium
since they suffer strong mass loss together with nucleo\-synthesis processes
\citep[see e.g.][]{busso1999}. Low-mass AGB stars (M $<$ 4 M$_{\sun}$)
can turn C-rich (i.e. C/O $>$ 1) because of the dredge-up of carbon from the bottom
of the convective envelope to the ste\-llar surface. The $s$-process a\-llows
the production of elements heavier than iron by $slow$ neutron-captures.  In
these stars, the $^{13}$C($\alpha$, n)$^{16}$O reaction is assumed to ope\-rate
as the main neutron source \citep[e.g.][]{abia2001}. On the other hand,
intermediate-mass AGB stars ( 4 $\leq$ M $\leq$  8 M$_{\sun}$) are O-rich
stars (C/O $<$ 1) because of the operation of hot bo\-ttom burning (HBB), which
burns carbon at the base of the convective envelope, thus preventing the
formation of a carbon star \citep{sackmann1992}.  In these carbon stars, the
$s$-process elements are expected to form mainly by the neutrons released by the
$^{22}$Ne($\alpha$, n)$^{25}$Mg reaction, in a higher neutron density
environment than in lower mass AGB stars \citep{garcia2006}. The relative
abundance of s-elements such as Rb to other neighboring ones such as Sr, Y, and Zr is
an indicator of the neutron density, namely a discrimi\-nant of the stellar mass
and the main neutron source at the $s$-process site
\citep{lambert1995,abia2001,garcia2006}.

Observationally, a low [Rb/Zr] ratio ($<$ 0) is found in low-mass AGB stars
\citep{plez1993,lambert1995,abia2001} while higher mass AGB stars display
[Rb/Zr] $>$ 0 \citep{garcia2006,garcia2007,garcia2009}.
\citet[][hereafter Paper I and Paper II, respectively]{garcia2006,garcia2009}
derived the Rb and Zr abundances in several Galactic and Magellanic Cloud
intermediate-mass AGB stars among a sample of OH/IR stars. The Rb
abundances and [Rb/Zr] ratios found in these objects re\-present a
challenge for theoretical AGB nucleosynthesis mo\-dels, which do not predict the
extreme Rb overabundances ([Rb/Fe] $\gtrsim$ 2 dex) and extraordinarily high
[Rb/Zr] ratios observed \citep{vanraai2012,karakas2012}. However, the Rb
abundance was derived from the resonant Rb I absorption line at 7800 \AA, using
hydrostatic model atmospheres. The Rb I line is probably affec\-ted by
contamination from one or more circumstellar (CS) components, as has already
been suggested by the detection of blue-shifted CS Rb I absorption lines in several
of these extreme O-rich AGB stars (see Paper I, II). 

In the present {\it Letter}, we explore for the first time the CS effects on the
Rb and Zr abundances derived in extreme O-rich AGB stars. To this end, we use more
realistic model atmospheres that include a gaseous CS envelope. The much lower
Rb abundances (and [Rb/Zr] ratios) derived here significantly alleviate the
actual mismatch between the AGB nucleosynthesis predictions and the optical
observations of intermediate-mass (4--8 M$_{\sun}$) Rb-rich AGB stars.


\section{The sample stars}

We selected a representative sample of five massive Rb-rich AGB stars from
Papers I and II: four Galactic stars and one LMC star, covering several OH expansion 
velocities, variability periods, and
metallicities
The rele\-vant
information for the sample stars is shown in Table \ref{table:1}. All stars
belong to the OH/IR class, with OH expansion velocities and variability periods
available in the literature. We chose those stars with high signal-to-noise (S/N
$\geq$ 30-50) optical spectra for reliable Rb and Zr abundance determinations.
The resolving power of the observations is R $=$$\lambda$/$\Delta$$\lambda$$=$
40,0000 to 50,000 for the Galatic stars and R$\sim$60,000 for the LMC star 
(see Papers I and II for further observational details). The sample stars display
a variety of Rb I 7800 \AA~line strengths and profiles. Two stars (IRAS
05098$-$6422 and IRAS 18429$-$1721) display weak non-shifted Rb I lines while
the other three stars (IRAS 06300$+$6058, IRAS 19059$-$2219, and IRAS
04498$-$6842) show a blue-shifted Rb I absorption feature that is indicative of
a CS envelope moving outwards in the line of sight\footnote{The
observed Rb I line in IRAS 06300$+$6058 is resolved in two components
(circumstellar and stellar).}. As was already shown in Papers I and II,
classical MARCS hydrostatic models are not able to  account for the blue-shifted
absorption feature. Furthermore, the Rb abundances derived in Papers I and II
with these hydrostatic model atmospheres are significantly larger than AGB
nucleosynthesis models predictions \citep[see e.g.][]{vanraai2012}. The most
remarkable case is the LMC star IRAS 04498$-$6842 with [Rb/M] =
$+$5.0\footnote{In Paper II, an average metallicity of [M/H]=$-$1.4, as
estimated from a few very weak metallic lines around the Rb I line, was used in
the synthesis. Here, however, we prefer to adopt the LMC average metallicity
([M/H] = $-$0.3) in the synthesis in order to be consistent with the original
metallicity of the atmosphere model.}.

\section{Abundance analysis using dynamical models} 

\subsection{Turbospectrum spectral synthesis code with extended atmospheres and
velocity fields} \label{turbo}
For a proper analysis of the circumstellar component around the
\mbox{Rb I} 7800 \AA\ line, we have modified the
v12.1 of the spectral synthesis code \textit{Turbospectrum}
\citep{alvarez1998,plez2012} to deal with  extended atmospheres and velocity
fields.  The Doppler effect due to the expanding envelope of the star is implemented
in \textit{Turbospectrum} by the modification of the routines that compute the line
intensities at the stellar surface. Concerning the radiative transfer, the source
function is assumed to be the same as computed in the static case without a wind
\citep[see][]{gustafsson2008}.  The sca\-ttering term of the source function
($\propto\sigma_{\lambda}${J$_{\lambda}$}) is not shifted to save
computing time. Scattering is only included for the continuum.
We carefully checked this approximation by comparing it
with Monte Carlo simulations (see Sect. \ref{MC}) because 
in cool CS envelopes photons are mostly scattered in resonance lines and are not
thermally emitted as we assume here.
In this way, the
sca\-ttering term is computed as in the static case  using the Feautrier method
\citep{nordlund1984,gustafsson2008}. The velocity field is taken into account
through a shift of the absorption coefficient $\kappa _{\lambda}$. The source
function is built using the static $\sigma_{\lambda}${J$_{\lambda}$ and the
shifted $\kappa _{\lambda}$B$_{\lambda}$. The emerging intensity is then
computed in the observer frame by a direct quadrature of the source function.
This algorithm is sensitive to numerical errors \citep{mihalas1986}. Therefore,
a high sampling of the optical depth and the wavelength points
is used to minimise errors in the opacity interpolation.

\subsection{Dynamical atmosphere models} \label{dm}  

The atmospheric parameters adopted for the individual analysis of each star are
taken from Papers I and II, using solar refe\-rence abundances by
\cite{grevesse2007}. Furthermore, our dynamical models are constructed from the
original MARCS hydrostatic atmosphere model structure, expanding the atmosphere
radius by the inclusion of a wind out to $\sim$ 5 stellar radii, with a
radial velo\-city field in spherical symmetry. The stellar radius $R_{*}$
is defined as the radius corresponding to $r(\tau_{Ross} = 1)$ in the MARCS
hydrostatic model, where r is the distance from the ste\-llar centre and
$\tau_{Ross}$ the Rosseland optical depth. The stellar wind is computed under
the assumptions of mass conservation (Eq. 1) and  radiative thermal equilibrium
(Eq. 2), following a classical $\beta-$velocity law (Eq. 3),

\begin{equation}
      \rho(r) = \frac{{\textit{\.M}}}{4 \pi r^2 v(r)} 
   \end{equation}
   
\begin{equation}
      r T^2 = constant  = r_{out}T^2_{out}
   \end{equation}
   
\begin{equation}
      v(r) = v_0+(v_{\infty}-v_0)(1-\frac{R_*}{r})^{\beta} \,,
   \end{equation}

\noindent 
where $\rho$($r$) is the density of the envelope at radius $r$, $\textit{\.M}$
is the mass-loss rate, and $v(r)$ is the velocity of the envelope. The velocity
$v(r)$ is calculated by means of Eq. 3, where $v_0$ is a reference velocity for
the onset of the wind and the $\beta$ exponent is an arbitrary free parameter. 
For the onset of the wind, we take $v_0 = v(R_*)$ and the extension of the
envelope begins from the outer radius of the hydrostatic model.  The envelope is
extended making use of Eq. 2, layer by layer, out to $r_{max}$. The distance
$r_{max}$ corresponds to the maximum radius in our calculations, with
temperature $T_{max}$ = 900 K, because \textit{Turbospectrum} cannot compute for
lower temperatures because of numerical reasons. For the mass-loss rate \textit{\.M}
and the $\beta$ exponent, we use values in the typical range of AGB stars: 
\textit{\.M} $\sim$ 10$^{-9}$--10$^{-4}$ M$_{\sun}$ yr$^{-1}$ in steps of 
factors of 10 and $\beta$ $\sim$ 0--1.6 in steps of 0.2. Finally, we assume
$v_{exp}(OH)$ as the terminal velocity $v_{\infty}$ (see Table \ref{table:1})
because the OH maser emission is found at very large distances of the central
star \citep[see e.g.][]{decin2010}. Figure \ref{models} shows examples of the
$\beta-$velocity law adopted in our dynamical models.

The resulting grid of synthetic spectra is compared to the observed spectra
in order to find the best fit to the Rb I line profile and the adjacent
pseudocontinuum. Figure \ref{RbK} shows several fits for the Rb I 7800 \AA\
and K I 7699 \AA\ lines (with similar atomic parameters) in IRAS 06300$+$6058 to
check the consistency of our models. In the left panel, we modify the
(\textit{\.M}, $\beta$) pair and the abundance of Rb until the best fit (in red)
to the observations (in black) is obtained. Remarkably, the best model
(\textit{\.M}, $\beta$) =  (1$\times$10$^{-7}$, 0.2) gives a Rb abundance of
+0.5, which is much lower than the hydrostatical abundance (see Table 1). Curiously,
the CS K I component is blue-shifted by $\sim$2-3 kms$^{-1}$ relative to that
in Rb I, suggesting a slightly higher velocity for the K I expanding gas.
Thus, we test the same models for the K I line fitting in the right panel but
varying the K abundance and the terminal velocity. Acceptable fits to
the K I line are obtained for solar K abundances and a slightly higher terminal
velocity of $\sim$15 kms$^{-1}$, as suggested by the blue-shifted CS K I line.
We note that increasing the K abundance may improve the fit a little bit. At
present we cannot discard some K production in these stars as the consequence of
proton-captures on Ar nuclei during HBB \citep{ventura2012}.

\subsection{Monte Carlo simulations} \label{MC}
As we already commented in Sect. \ref{turbo}, to check if our neglect of line
scattering in \textit{Turbospectrum} does not affect the interpretation of the
observed spectra, we performed complementary Monte Carlo simulations (MCS) that
only take into account photon sca\-ttering for the radiative transfer.
Three-dimensional MCS were carried out throwing a large number of photons
($\sim$10$^6$ for accurate statistics) through the expanding envelope of a star,
taking the initial wavelength, optical depth at which the photon will be
scattered $\tau_{\lambda}$, and throwing direction of the incident photon
as random numbers.  The photon of wavelength $\lambda$ has an absorption
probability 1$-$e$^{-\tau_{\lambda}}$. The random $\tau_{\lambda}$ at which the
photon will be scattered follows uniform statistics, i.e. the same
number of photons are thrown for each wavelength in all directions. Thus,  we
study the emergent photon distribution (spectrum) obtained for the CS component
around  the K I 7699 \AA\ line. The results can be extrapolated to the CS
component of the Rb I 7800 \AA\ line, taking into account that the density of K
is $\sim$ 300 times the Rb density for solar composition. 
As we only want to study how the line profile
changes qualitatively with a very simple code
the parameters describing the envelope are assumed constant 
in each simulation: temperature, T;
(K); density of K I atoms, n (cm$^{-3}$); expansion velocity, v$_{exp}$ (km
s$^{-1}$); and maximum envelope size, R$_{max}$ (cm, 1
AU$\sim$1.5$\times$10$^{10}$ cm),   We performed 216 simulations in
total, covering the following space of parameters: T =  10, 100, 1000 K; n  =
1$\times10^{-6}$,  1$\times10^{-5}$, 1$\times10^{-4}$, 1$\times10^{-3}$
cm$^{-3}$;  v$_{exp}$    =  5, 10, 15 km s$^{-1}$ and R$_{max}$  = 
1$\times10^{13}$, 1$\times10^{14}$, 1$\times10^{15}$, 1$\times10^{16}$,
1$\times10^{17}$, 1$\times10^{18}$ cm. A re\-presentative subset of the results
obtained for T = 1000 K - the coolest temperature that
\textit{Turbospectrum} is able to compute - and v$_{exp}$ = 10 km s$^{-1}$  are
shown in Fig. \ref{MCresults}. For other envelope temperatures or expansion
velocities, the resul\-ting line profiles are very similar. In short, the most
frequent line profile obtained from MCS is a blue-shifted absorption feature
with a weak and broader emission component. The envelope parameters of the
simulations displaying a blue-shifted absorption feature with weak emission are
within the range of the dynamical atmosphere models used in
\textit{Turbospectrum}, which ends at R$_{max}$ = 1.6$\times10^{14}$ cm.
For example, for R$_{max}$ = 1.6$\times10^{14}$ cm some weak emission is
obtained in two MCS only, corresponding to a higher density of n =
1$\times10^{-3}$ cm$^{-3}$. PCygni profiles, i.e. with a strong emission
component, are less frequent (14\%) and they are obtained only at very high
densities and/or large envelope sizes. In the rest (30\%) of the MCS, the spectrum
is flat and no absorption and/or emission line is found in the
distribution of the emergent photons for very low density (n $\leq$
1$\times10^{-5}$ cm$^{-3}$) and small envelope size (R$_{max}$ $<$ 10$^{15}$ cm)
values (Fig. \ref{MCresults}).

The optical observations of Rb in massive Galactic AGB stars tell us that a
non-shifted (purely photospheric) or a blue-shifted Rb I (with a CS component)
line are by far the most frequent cases  (see Paper I). There is only one star
IRAS 12377$-$6102 with a clear PCygni Rb I profile; perhaps the star
IRAS 18050$-$2213 also shows some weak emission. These two stars are among the
reddest Rb-rich AGB stars, also showing very high OH expansion velocities
($\sim$20 km s$^{-1}$), and they are presumably the most evolved and extreme
stars. Indeed, \citet{justtanont2013} estimate envelope sizes from
$\sim$9x10$^{15}$ cm to 2.5x10$^{16}$ cm in a few Galactic OH/IR stars
(massive AGBs) from Herschel observations. Only one star in this work, WX
Psc, has an optical counterpart. Although the S/N around 7800 \AA\ is quite
low, a strong Rb I absorption line is clearly detected in the WX Psc optical
spectrum (Paper I). Interestingly, the estimated mass-loss rate for WX Psc
($\sim$10$^{-5}$ M$_{\odot}$ yr$^{-1}$) is lower by one order of
magnitude than that for the more obscured stars. Furthermore, Betelgeuse
observations \citep{plez2002} show that a mass-loss rate of $\sim$ 10$^{-6}$
M$_{\odot}$ yr$^{-1}$ results in a double K I 7699 \AA\ line with emission in
the center thanks to the photon sca\-ttering but this emission is feeble in
the integrated star spectrum and the circumstellar absorption dominates. Taking
into account the results of the MCS and the cited observational constraints, we
conclude that \textit{Turbospectrum} (Sect. \ref{turbo}), provides a reasonable
approximation for the stars studied here. 

\section{Results and discussion}

The parameters of the dynamical atmosphere models providing the best fit to the
observations and the derived Rb and Zr abundances ([Rb/M]$_{dyn}$ and
[Zr/M]$_{dyn}$) are shown in Table \ref{table:2}. The Rb abundances
([Rb/M]$^{ref}_{static}$) as determined in Papers I and II from hydrostatic
models are also shown for comparison. We note, however, that Papers I and II used
solar abundances from \cite{grevesse1998} while we assume here the most recent
solar composition by \cite{grevesse2007}. For this reason, Table \ref{table:2}
also lists the re-derived Rb abundances ([Rb/M]$_{static}$) using the
hydrostatic models with our adopted solar abundances. Our static Rb abundances
agree well, within the errors, with those previously derived in Papers I
and II.

\begin{table*}
\tiny
\begin{center}
\caption{Atmosphere parameters and abundances derived using dynamical models vs. hydrostatic models.}
\label{table:2}
\centering
\begin{tabular}{c c c c c c| c c c c}
\noalign{\smallskip}                                                                                 
\hline\hline  
\noalign{\smallskip}
IRAS name&$T\rm{_{eff}}$ (K)&$\rm \log$ g& $\beta$& \textit{\.M} (M$_{\sun}$ yr$^{-1}$)&$v$ (km
s$^{-1}$)&[Rb/M]$^{ref^a}_{static}$&[Rb/M]$_{static}$$^b$&[Rb/M]$_{dyn}$$^b$&[Zr/M]$_{dyn}$$^b$\\
\hline
\noalign{\smallskip}
\multicolumn{10}{c}{Galactic stars}\\
 \hline
 \noalign{\smallskip}

 05098$-$6422  &3000  &  $-$0.5 & 1.0& 1.0 $\times$ 10$^{-8}$    &6 & 0.1    &0.0 $\pm$ 0.4&0.0 $\pm$ 0.4  &$\leq$ 0.3 $\pm$ 0.3        \\
 06300$+$6058  &3000  &  $-$0.5 & 0.2& 1.0 $\times$ 10$^{-7}$    &12& 1.6    &1.9 $\pm$ 0.4&0.5 $\pm$ 0.7  &$\leq$ 0.1 $\pm$ 0.3    \\
 18429$-$1721  &3000  &  $-$0.5 & 1.0& 1.0 $\times$ 10$^{-8}$    &7& 1.2    &1.2 $\pm$ 0.4&1.0 $\pm$ 0.4  &$\leq$ 0.3 $\pm$ 0.3     \\
 19059$-$2219  &3000  &  $-$0.5 & 0.4& 1.0 $\times$ 10$^{-7}$    &13& 2.3/2.6&2.4 $\pm$ 0.4&0.8 $\pm$ 0.7  &$\leq$ 0.3 $\pm$ 0.3         \\

\noalign{\smallskip}
\hline
\multicolumn{10}{c}{LMC star}\\
\hline\noalign{\smallskip}
 04498$-$6842  &3400  &  0.0    & 1.0& 1.0 $\times$ 10$^{-7}$   &13& 3.9$^{c}$    &3.3 $\pm$ 0.4&1.5 $\pm$ 0.7   &$\leq$ 0.3 $\pm$ 0.3     \\
\hline     
\end{tabular}

\tablefoot{
\\
\tablefoottext{a}{See Paper I and II.} 
\tablefoottext{b}{The uncertainties represent the formal errors due to
the sensitivity of the derived abundances to slight changes in the model
atmosphere parameters ($\Delta${\it T}$_{eff}$=$\pm$100 K,
$\Delta$[M/H]=$\pm$0.3, $\Delta$$\xi$=$\pm$1 kms$^{-1}$, $\Delta$log {\it
g}=$+$0.5, $\Delta$FWHM=50 m\AA, $\Delta$$\beta$=0.2, 
$\Delta$log(\.M/M$_{\odot}$yr$^{-1}$)=1.0}) for each star.
\tablefoottext{c}{We scale the Rb overabundance derived by Paper II, [Rb/M] =
+5.0, to the adopted LMC metallicity [M/H] = $-$0.3.}}
\end{center}
\end{table*}

\begin{figure}
   \includegraphics[width=7cm,angle=-90]{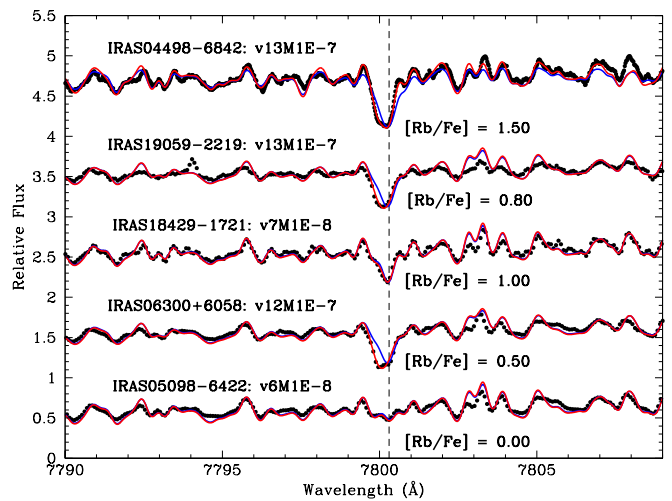}
      \caption{Rb abundances derived in the sample stars using dynamical models.
      The location of the Rb I stellar line is indicated by a dashed line.
      Dynamical models providing the best fits to the observations (black dots)
      are indicated by a red line. Hydrostatic models are also shown for
      comparison (blue lines). The expansion velocity and the mass-loss rate
      adopted in the models also are indicated for each star.}
         \label{Rb}
   \end{figure}

Interestingly, the new Rb abundances (in the range [Rb/M]$\sim$0.0$-$1.5 dex;
Table \ref{table:2}) derived from our dynamical models display a dramatic
decrease of 1.4 dex to 1.8 dex with respect to the static case (Table
\ref{table:2}) in the more extreme stars, namely those stars showing the highest
expansion velocities and mass-loss rates such as IRAS 06300$+$6058, IRAS
19059$-$2219 and IRAS 04498$-$6842. However, for the less extreme stars with a
lower expansion velocity and mass-loss rate (IRAS 05098$-$6422 and IRAS
18429$-$1721), the Rb abundances obtained from dynamical models remain close, within 0.2 dex,
to those from hydrostatic models. In Fig. \ref{Rb}, we
display the observed Rb I line profiles in our O-rich AGB sample (black dots)
together with the best synthetic spectra as obtained from the new dynamical
models (red lines) versus the static ones (blue lines). Our dynamical atmosphere
models reproduce the observed Rb I line profiles (photospheric and
circumstellar components) very well, much better than the classical
hydrostatic models. The 6474 \AA\ ZrO bandhead, because it is formed deeper in the
atmosphere, is less affected than the Rb I line, and the Zr abundances
(nearly-solar; see Fig. \ref{Zr}) derived from dynamical models are
similar to those obtained with the hydrostatic models (see Papers I and
II).   

Remarkably, we find that relatively low mass-loss rates
($\sim$10$^{-7}$$-$10$^{-8}$ M$_{\odot}$ yr$^{-1}$) give superior fits to the
observed Rb I line profiles. Higher mass-loss rates of $\geq$10$^{-6}$
M$_{\odot}$ yr$^{-1}$ give Rb I absorption lines that are too strong even for solar Rb
abundances (see Fig. \ref{RbK}). According to \cite{justtanont2013}, the
envelope size of $\sim$1x10$^{16}$ cm and the mass-loss rate of $\sim$10$^{-4}$
M$_{\odot}$ yr$^{-1}$ estimated in optically obscured OH/IR AGB stars can be
taken as upper limits for massive AGB stars with optical counterparts; i.e.
like our sample stars with useful spectra around the Rb I line because
the heavily obscured OH/IR massive AGB stars studied by \cite{justtanont2013}
have already entered the superwind phase. Our best fits for the Rb I
7800 \AA\ line profiles in optically bright massive AGB stars are those with
mass-loss rates lower than $\sim$10$^{-6}$ M$_{\odot}$ yr$^{-1}$. Thus, we
conclude that the Rb I line profiles observed in massive AGB stars are
consistent with these stars still experiencing relatively low mass-loss rates
just before the superwind phase when the Rb density and the envelope size
are not very high and this is supported by the MCS presented above (Sect.
3.3).

Standard nucleosynthesis models for intermediate-mass AGB stars show that the
predicted Rb abundances range from [Rb/M]$\sim$0.0 up to 1.44 dex, depending on
the progenitor mass and metallicity \citep[see][]{vanraai2012}; the predicted Rb
production increases with increasing stellar mass and decreasing metallicity.
Maximum [Rb/M] overabundances of 1.04 and 1.44 are found for a solar metallicity
6.5 M$_{\sun}$ star and for a LMC metallicity 6 M$_{\sun}$ star, respectively
\citep{vanraai2012}. More recently, \cite{karakas2012} have delayed the
beginning of the superwind phase in solar metallicity nucleosynthesis models of
massive AGB stars. These models produce more Rb than in the standard van Raai et
al. models because the star experiences more thermal pulses before the superwind
phase at the very end of the AGB. The maximum Rb production ([Rb/M]=1.34 dex) is
predicted to occur for the 6 M$_{\sun}$ case. By considering the error bars in
the spectroscopic analysis (see Table \ref{table:2}) and the theoretical
uncertainties \citep[see e.g.][]{vanraai2012,karakas2012}, the Rb abundances
are now in fair agreement with the massive AGB nucleosynthesis models, both standard 
and with delayed superwinds. The nearly-solar derived Zr
abundances in IRAS 05098$-$6422,  IRAS 06300$+$6058, and IRAS 19059$-$2219
(Table \ref{table:2}) translate into [Rb/Zr] ratios of $-$0.3, 0.4, and 0.5,
respectively, which agree quite well with the theoretical predictions (-0.2
$\leq$ [Rb/Zr] $<$0.6). However, the [Rb/Zr] ratios in IRAS 18429-1721 (0.7) and
IRAS 04498-6842 (1.2) are still higher than predicted. As already pointed out in
the literature \citep[see e.g.][]{vanraai2012}, a possible solution to this
observational problem is that gaseous Zr, with a condensation temperature (1741
K) higher than that for Rb (800 K), condensates into dust grains, producing the
apparent Zr underabundance that we measure from the ZrO bandheads.

In summary, the Rb abundances and [Rb/Zr] ratios derived here significantly
resolve the problem of the present mismatch between the observations of
massive (4--8 M$_{\sun}$) Rb-rich AGB stars and the theoretical predictions. In
the near future, we plan to carry out a chemical analysis based on these new
dynamical models for all the Rb-rich AGB stars already studied in Papers I and II.
This undoubtedly will help us to constrain the actual nucleosynthesis models for
the more massive AGB stars. 

\begin{acknowledgements}
O.Z., D.A.G.H. and A.M. acknowledge support provided by the Spanish Ministry of
Economy and Competitiveness under grant \emph{AYA-2011-27754}. 
\end{acknowledgements}

\bibliographystyle{aa} 
\bibliography{biblio.bib} 

\Online
\begin{appendix}

\section{Table A.1 and Figures A.1, A.2, and A.3}

\begin{table*}
\begin{center} 
\caption{The sample stars}             
\label{table:1}      
\begin{tabular}{c c c c}        
\hline\hline                 
\noalign{\smallskip}
IRAS name &            SpType & $v_{exp}(OH)$ (km s$^{-1}$)& Period (days)   \\
\hline\noalign{\smallskip}
\multicolumn{4}{c}{Galactic stars}\\
\hline \noalign{\smallskip}    
   05098$-$6422 &M7e &6.0           &  394  \\   
   06300$+$6058 &M7+ &12.3          &  440  \\
   18429$-$1721 &M9  &6.9           &  1508  \\      
   19059$-$2219 &M8  &13.3          &  510    \\           
\hline\noalign{\smallskip}
\multicolumn{4}{c}{LMC star}\\
\hline \noalign{\smallskip}                               
   04498$-$6842 &M   &13.0          &  1292  \\
\hline                                   
\end{tabular}
\end{center}
\scriptsize{}.
\end{table*}

\begin{figure*}
   \centering
   \includegraphics[width=7.0cm]{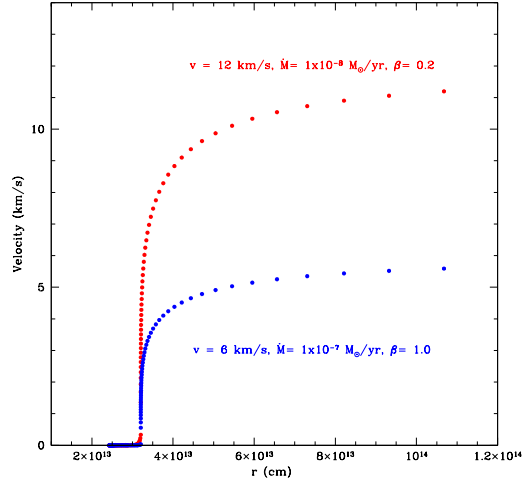}
      \caption{Velocity law (velocity vs. distance from the star) in two of our
AGB wind models. Both models are based on the MARCS hydrostatic model with
T$_{eff}$ = 3500 K, log$g$ = $-$0.5, and solar chemical composition. The red
model is computed with velocity at infinite of 12 km s$^{-1}$, mass-loss rate of
10$^{-8}$ M$_{\sun}$ yr$^{-1}$, and $\beta$ exponent of 0.2. The blue model has a
velocity at infinite of 6 km s$^{-1}$, mass-loss rate of 10$^{-7}$ M$_{\sun}$
yr$^{-1}$, and $\beta$ exponent of 1.0. We note that a smaller step is taken where
the velocity gradient is steepest.
              }
         \label{models}
   \end{figure*}

\begin{figure*}
   \centering
   \includegraphics[width=7cm,angle=0]{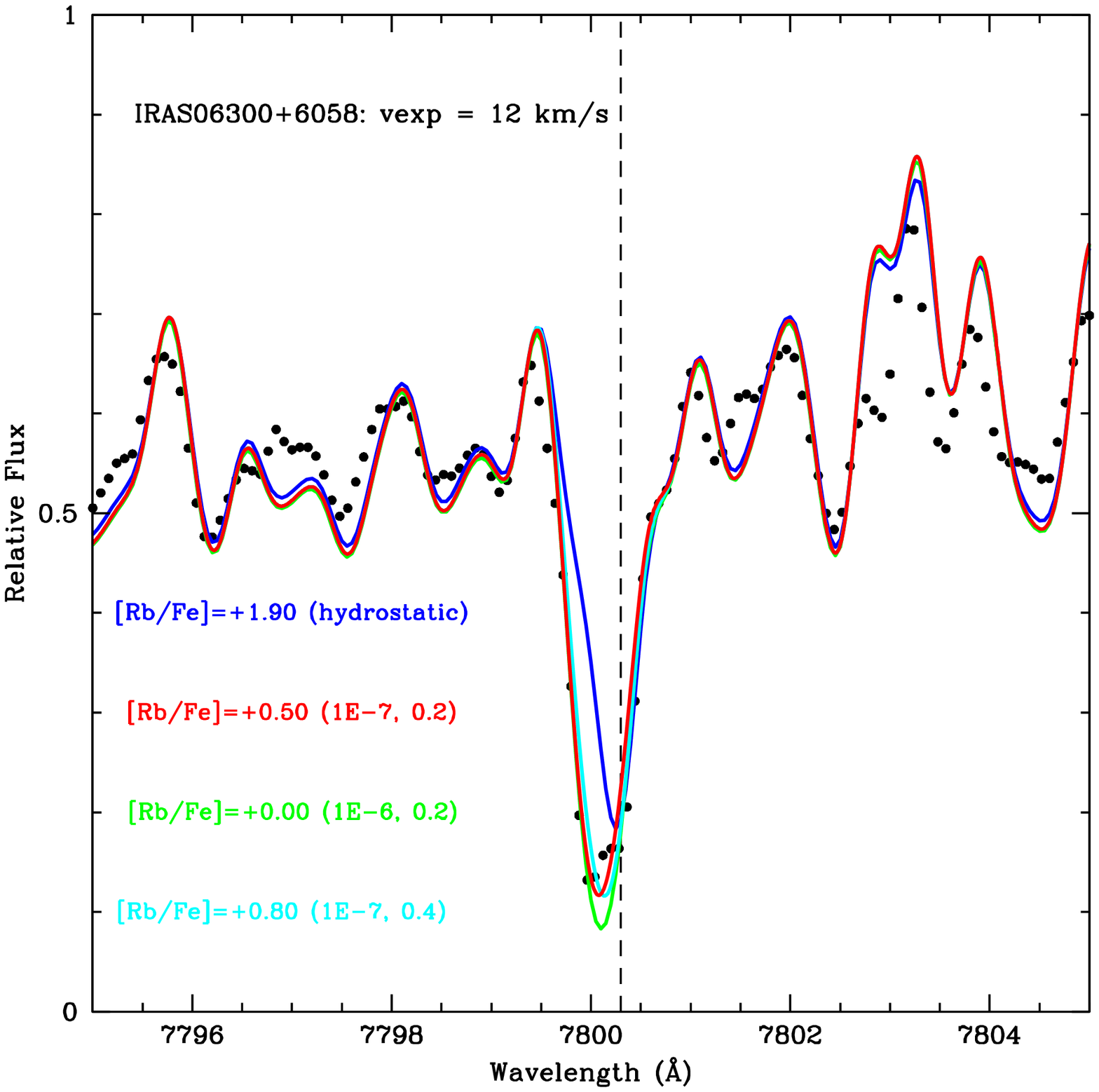}
    \includegraphics[width=7cm,angle=0]{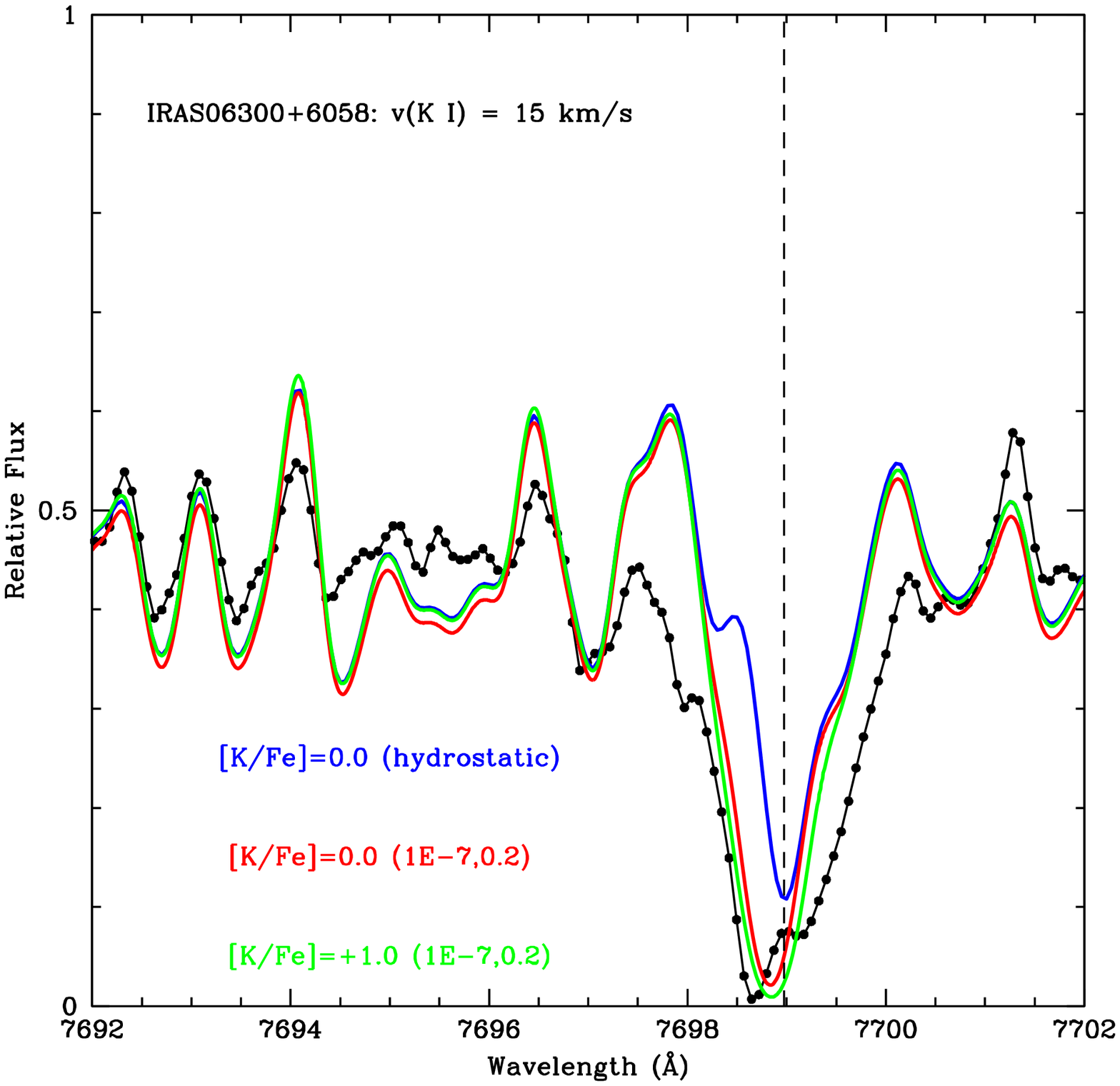} 
      \caption{Rb I (left panel) and K I (right panel) line profiles obtained
using dynamical models given by the (\textit{\.M}, $\beta$) = 
(1$\times$10$^{-7}$, 0.2) pair (see Sect. \ref{dm}) for the star IRAS
06300$+$6058. The hydrostatic model (blue line) is also shown for comparison.
The colours have similar meaning in the two panels but in the case of K I, all
models were computed with a sligthly higher terminal velocity of 15 kms$^{-1}$,
as suggested by the blue-shift of the K I line. We note that the red wing of the K
I profile is not fully reproduced because of the presence of at least one
interstellar component that is not seen in Rb I.}
         \label{RbK}
   \end{figure*}

\begin{figure*}
   \centering
   \includegraphics[width=7cm,angle=0]{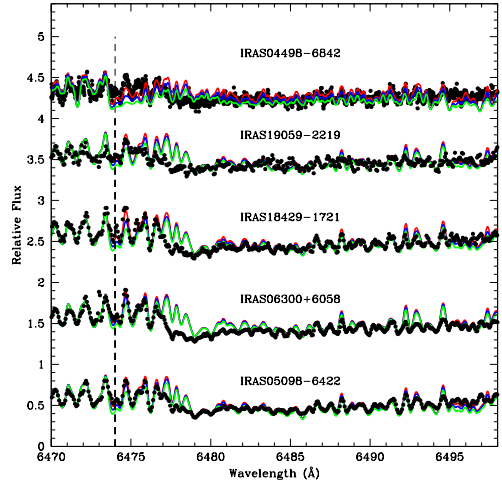}
      \caption{High-resolution optical spectra (in black) and dynamical models that provide the best fits (in red) in the ZrO 6474 \AA{} region for our
      sample stars. Synthetic spectra obtained for Zr abundances shifted +0.25 dex (in blue) and +0.50 dex (in green) from the adopted values are also shown.}
         \label{Zr}
   \end{figure*}

\section{Monte Carlo simulations; Fig. B.1}

 \begin{figure*}
\centering
\includegraphics[width=\textwidth,angle=0]{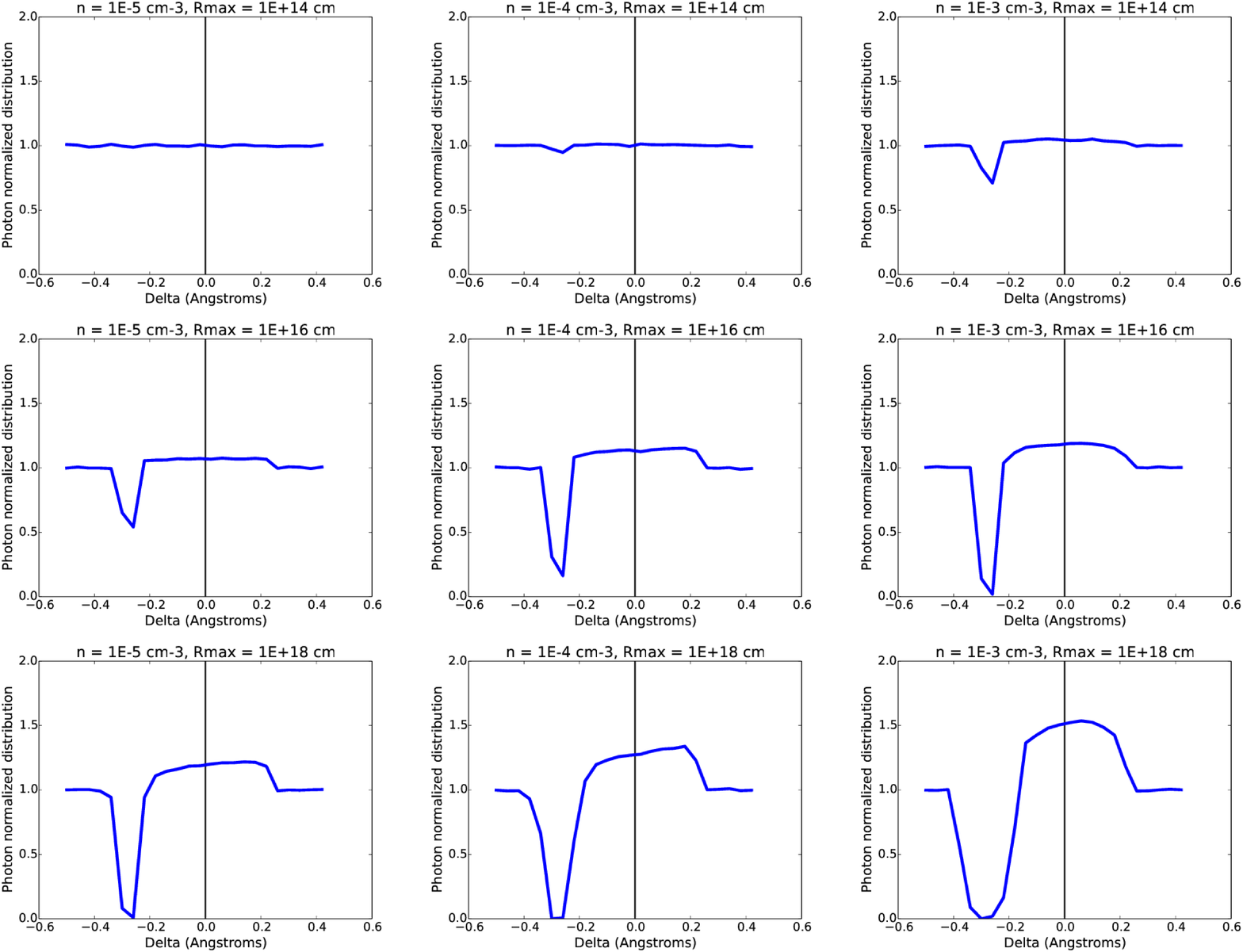}
 \caption{Illustrative example of nine Monte Carlo simulations carried out with
the envelope parameters T = 1000 K and $v_{exp}$ = 10 km s$^{-1}$. The density
(n, in cm$^{-3}$) and the envelope size (R$_{max}$, in cm) increase from left
to right and from top to bottom, respectively. We show the displacement (in
Angstroms) from the rest wavelength (Delta=$\lambda-\lambda_0$, where
$\lambda_0$ is indicated by a vertical line) to the emergent wavelength
distribution.}
 \label{MCresults}
\end{figure*} 

\end{appendix}

\end{document}